# Structure and dynamics of spray detonation in *n*-heptane droplet / vapor / air mixtures

Qingyang Meng[1], Majie Zhao[2], Yong Xu[2], Liangqi Zhang[3], and Huangwei Zhang[2,*]

[1] National University of Singapore (Chongqing) Research Institute, Liangjiang New Area, Chongqing 401123, People's Republic of China
[2] Department of Mechanical Engineering, National University of Singapore, 9 Engineering Drive 1, Singapore 117576, Republic of Singapore
[3] Department of Engineering Mechanics, College of Aerospace Engineering, Chongqing University, Chongqing 400044, People's Republic of China

**Abstract**

Spray detonation in *n*-heptane droplet / vapour / air mixtures is simulated using Eulerian–Lagrangian method. Two-dimensional configuration is considered, and the effects of droplet diameter and liquid equivalence ratio on detonation propagation, structure, and dynamics are investigated. The results show that the average detonation propagation speed first increases and then decreases as liquid equivalence ratio changes, and the speed peaks at higher liquid equivalence ratio for larger droplets. The triple points / transverse detonations vaporize or aerodynamically expel the droplets from their trajectories, resulting in non-uniform distributions of fuel vapour and reaction zones behind the detonation. In addition, droplet dispersion distance in the post-detonation area increases for larger droplets due to lower evaporation. Moreover, small droplets generally lead to higher detonated *n*-heptane fraction, and fuel detonative combustion directly affects the variations of detonated fuel fraction. For larger droplets, V-shaped dependence on liquid equivalance ratio is seen for large droplets, dominated by variations of post-detonation deflagration. It is found that spray detonation structure is signifciantly infuenced by liquid fuel equivalance ratio and droplet diameter. The dependence of key locations in spray detonation structure on liquid fuel properties is also evaluated, e.g., reaction front and sonic plane. Furthermore, the leading shock Mach number slightly decreases with droplet size. When the liquid equivalence ratio is high, spray detonation exhibits pronounced unsteadiness, such as instantaneous or complete extinction. Either extinction is caused by strong heat absorption of evaporating droplets behind the shock. Moreover, localized detonative spot is observed due to the compression of multiple transverse shocks.

---

[*] Corresponding author. E-mail: huangwei.zhang@nus.edu.sg. Tel: +65 6516 2557.

# 1. Introduction

Detonative combustion technology has great potential to revolutionize the existing propulsion system, because of its numerous advantages, e.g., pressure gain, large thrust-weight ratio, and high thermal efficiency [1-3]. Most previous detonation engine investigations (e.g., rotating detonation engine, RDE, or pulse detonation engine, PDE) consider gaseous fuels, such as hydrogen [4-6]. However, utilization of liquid fuels is a significant step towards commercializing detonation engines, since they have high energy density and wide availability [7].

Liquid hydrocarbons have been investigated in a range of laboratory-scale detonation engines or idealized equivalent configurations. For instance, Fan et al. [8] studied a model PDE with $C_8H_{16}$ / air mixture and successfully achieved pulsed detonations up to 36 Hz in tubes of different lengths. Moreover, experiments about liquid hydrocarbon (kerosene, gasoline, etc.) RDE were also conducted. The results demonstrate the significance of burner configuration [9, 10] and pre-vaporization of liquid fuels [11] to achieve a continuous detonation. More recently, Liu et al. [12] experimentally tested kerosene atomization and cold spray mixing characteristics with a linearized combustor. Through considering the preheated air and liquid *n*-heptane, Jin et al. [13] revealed the chemical structures in the fuel refill zone in a modelled RDE combustor and examined the low-temperature chemistry effects on propulsion indices. Nonetheless, our fundamental understanding about spray detonations is still lacking, due to the intrinsic complexities of two-phase reaction systems and limitations of the existing diagnostic techniques.

How dispersed fuel droplets affect the detonation structure is one of the important questions in spray detonations. Ragland et al. [14] measured dimethyl-cyclohexanes detonation structures and found that the reaction zone thickness is several orders of magnitude larger than that of a gaseous detonation. They further revealed the droplet dynamics within the reaction zone, e.g., breakup, localized explosion and burning. Borisov et al. [15] found that the reaction zone would be lengthened without droplet breakup and the detonation cannot sustain. This is because droplet breakup is accompanied by local explosion in the wake of the leading shock, which is beneficial to facilitate detonation sustainability [16]. Local explosion induced by dispersed droplets is also observed by Kauffman and Nicholls [17], which takes place in the wake of the partially shattered droplets after ignition delay.

Moreover, fuel droplet properties (e.g., size, polydispersity, or loading) exhibit various influences on detonation propagation behaviors. Many studies have shown that the propagation speed of spray detonation are close to that of gaseous detonation when fine droplets (decane, hexadecane, propane, etc.) are loaded [18, 19]. However, relatively large droplets ( > 50 μm) may result in velocity deficit [20]. This is also observed in spray RDE [21-25] and PDE studies [26, 27]. Eidelman and Burcat [20, 28] attributed detonation speed deficit in large-droplet detonations to increased reaction zone thickness. Furthermore, Dabora et al. [29] focused on the detonation speed of $C_{10}H_{20}$ spray detonations considering different droplet size distributions, e.g., polydisperse, monodisperse, and film. They obtained about 30% velocity deficit compared to the theoretical value with monodisperse sprays with droplet size of about 940 μm. Shen et al. [30], Benmahammed et al. [31], Kadosh and Michaels [27] tested the effects of total equivalence ratio (i.e., from both gas and liquid phase) on spray detonation speed. They obtained the U-shaped relation between the detonation speed and equivalence ratio, and the velocity deficit from their measurements is $9-20\%$. Recently, Jourdaine et al. [32] investigated the effects of droplet size distributions on *n*-heptane / air detonation speed based on a sampling model using Eulerian−Eulerian method. They found that the biggest velocity deficit (~ 4.8%) was obtained in case of 15 samples which was the optimal approximation of the practical droplet size distribution in their study, and about 2.0% velocity deficit is obtained in case of 1 sample, i.e., monodispersed droplet case.

In spite of the aforesaid research progress, spray detonation structure and unsteady phenomena under different liquid fuel conditions are still not well understood, which however are of primary importance for designing effective and stable detonation propulsion system with liquid fuels. In the present work, *n*-heptane is considered, because it is a major component of many real fuels, e.g., gasoline and jet fuel. We aim to investigate the influences of liquid *n*-heptane spray properties (i.e., initial droplet diameter and liquid equivalence ratio) on structure and dynamics of spray detonations. Our research objectives include: (1) structure of shock and reaction fronts in two-phase *n*-heptane / air detonations, (2) detonation / deflagration reaction zone distribution and detonated fuel fraction, and (3) transient spray detonation phenomena (e.g., decoupling and re-initiation) and underpinning mechanisms. The manuscript is organized as below. In Section 2, the computational method is introduced and physical model is given in Section 3. Results and discussion are presented in Section 4, followed by conclusions in Section 5.

## 2. Mathematical model

The Eulerian−Lagrangian method is employed to simulate detonation wave propagation in two-phase *n*-heptane / air mixtures. *n*-Heptane sprays are tracked with the Lagrangian method. Inter-droplet interactions (such as collision or coalescence) are neglected since droplet volume fraction is less than 0.1% [33]. Droplet breakup by the aerodynamic force is not considered due to the smallness of the droplets. The droplet temperature is assumed to be uniform because of small Biot number. With above assumptions, the equations of gas and liquid phases are outlined as below.

## 2.1 Governing equation

The equations of mass, momentum, energy, and species mass fraction are solved for the gas phase

$$\frac{\partial \rho}{\partial t} + \nabla \cdot [\rho \mathbf{u}] = S_m, \tag{1}$$

$$\frac{\partial (\rho \mathbf{u})}{\partial t} + \nabla \cdot [\mathbf{u}(\rho \mathbf{u})] + \nabla p + \nabla \cdot \mathbf{T} = \mathbf{S_F}, \tag{2}$$

$$\frac{\partial (\rho E)}{\partial t} + \nabla \cdot [\mathbf{u}(\rho E)] + \nabla \cdot [\mathbf{u}p] + \nabla \cdot [\mathbf{T} \cdot \mathbf{u}] + \nabla \cdot \mathbf{q} = \dot{\omega}_T + S_e, \tag{3}$$

$$\frac{\partial (\rho Y_m)}{\partial t} + \nabla \cdot [\mathbf{u}(\rho Y_m)] + \nabla \cdot \mathbf{s_m} = \dot{\omega}_m + S_{Y_m}. \tag{4}$$

Here $t$ is time and $\nabla \cdot (\cdot)$ is the divergence operator. $\rho$ is the gas density, and $\mathbf{u}$ is the gas velocity vector. $p$ is the pressure updated from the idea gas equation of state, i.e., $p = \rho RT$., in which $T$ is the gas temperature and $R$ is the specific gas constant. $Y_m$ is the mass fraction of $m$-th species. $E \equiv e + |\mathbf{u}|^2/2$ is the total non-chemical energy, and $e$ is the specific internal energy. The source terms in Eqs. (1)−(4), i.e., $S_m$, $\mathbf{S_F}$, $S_e$ and $S_{Y_m}$, denote the exchanges of mass, momentum, energy and species between the gas and liquid phases, and their expressions are given in Eqs. (9)−(12).

The viscous stress tensor $\mathbf{T}$ in Eq. (2) is modelled as $\mathbf{T} = -2\mu \text{dev}(\mathbf{D})$. Here $\mu$ is the dynamic viscosity of the gas mixture and estimated with the Sutherland's law. Moreover, $\text{dev}(\mathbf{D}) \equiv \mathbf{D} - \text{tr}(\mathbf{D})\mathbf{I}/3$ is the deviatoric component of the deformation gradient tensor $\mathbf{D}$, which is $\mathbf{D} \equiv [\nabla \mathbf{u} + (\nabla \mathbf{u})^T]/2$. $\mathbf{I}$ is the unit tensor. In addition, the diffusive heat flux $\mathbf{q}$ in Eq. (3) is modelled with Fourier's law, i.e., $\mathbf{q} = -k\nabla T$. The thermal conductivity $k$ is calculated using the Eucken approximation [34]. In Eq. (4), $\mathbf{s_m} = -D_m \nabla(\rho Y_m)$ is the species mass flux. The mass diffusivity $D_m$ is derived from the heat diffusivity $D_m = \alpha = k/\rho c_p$ with unity Lewis number assumption. $c_p$ is the heat capacity at constant pressure. Moreover, $\dot{\omega}_m$ is the reaction rate of $m$-th species by all reactions, and $\dot{\omega}_T$ in Eq. (3) accounts for combustion heat release rate.

For the liquid phase, computational parcel method is used to group the liquid fuel droplets with identical properties (e.g., size, velocity, and temperature). The evolutions of droplet mass, momentum, and energy in a parcel are governed respectively by

$$\frac{dm_d}{dt} = -\dot{m}_d, \tag{5}$$

$$\frac{d\mathbf{u}_d}{dt} = \frac{\mathbf{F}_d}{m_d}, \tag{6}$$

$$c_{p,d}\frac{dT_d}{dt} = \frac{\dot{Q}_c + \dot{Q}_{lat}}{m_d}. \tag{7}$$

Here $m_d = \pi \rho_d d^3/6$ is the droplet mass, where $\rho_d$ and $d$ are the droplet material density and diameter, respectively. $\mathbf{u}_d$ is the droplet velocity vector, $c_{p,d}$ is the droplet heat capacity, and $T_d$ is the droplet temperature.

The droplet evaporation rate, $\dot{m}_d$, is calculated with the Abramzon and Sirignano model [35], i.e.,

$$\dot{m}_d = \pi d \rho_f D_f \widetilde{Sh} \ln(1 + B_M), \tag{8}$$

where $\rho_f = p_S MW_m/RT_S$ and $D_f = 3.6059 \times 10^{-3} \cdot (1.8T_s)^{1.75} \cdot (\alpha/p\beta)$ are the density and mass diffusivity at the film [35], respectively. $\alpha$ and $\beta$ are the constants related to specific species ($\alpha = 0.214$ and $\beta = 2.83$ for $n$-heptane) [36]. The surface vapor pressure $p_S$ is estimated from $p_S = p \cdot \exp(c_1 + c_2/T_s + c_3 \ln T_s + c_4 T_s^{c_5})$. For $n$-heptane, the constants, $c_1 - c_5$, are 87.829, -6996.4, -9.8802, 7.21×10⁻⁶ and 2.0, respectively [37]. Moreover, the droplet surface temperature $T_S$ is estimated from $T_S = (T + 2T_d)/3$ [35].

The modified Sherwood number $\widetilde{Sh}$ in Eq. (8) is calculated as $\widetilde{Sh} = 2 + [(1 + Re_d Sc)^{1/3} \max(1, Re_d)^{0.077} - 1]/F(B_M)$, with the Schmidt number being $Sc = 1.0$. The function $F(\vartheta) = (1+\vartheta)^{0.7} \ln(1+\vartheta)/\vartheta$ is introduced to consider the variation of the film thickness due to Stefan flow effects [35]. The Spalding mass transfer number is $B_M \equiv (Y_{Fs} - Y_{F\infty})/(1 - Y_{Fs})$, in which $Y_{Fs}$ and $Y_{F\infty}$ are the fuel vapor mass fractions at the droplet surface and ambient mixture, respectively. The former is calculated from $Y_{Fs} = MW_d X_s / [MW_d X_s + MW_{ed}(1 - X_s)]$, where $MW_d$ is the molecular weight of the vapor, $MW_{ed}$ is the averaged molecular weight of the mixture excluding the fuel vapor, and $X_S = X_m p_{sat}/p$ is the mole fraction of the vapor at the droplet surface. Here $X_m$ is the molar fraction of the condensed species in the gas phase. $p_{sat}$ is the saturation pressure and calculated based on Raoult's Law [38], i.e., $p_{sat} = p \cdot exp(c_1 + c_2/T_d + c_3 \ln T_d + c_4 T_d^{c_5})$. The constants, $c_1 - c_5$, take the same values for the surface vapor pressure $p_S$.

The Stokes drag in Eq. (6) is modelled as $\mathbf{F}_d = (18\mu/\rho_d d^2)(C_d Re_d/24) m_d (\mathbf{u} - \mathbf{u}_d)$. $C_d$ is the drag coefficient and estimated using the Schiller and Naumann model [39], and $Re_d \equiv \rho d |\mathbf{u}_d - \mathbf{u}|/\mu$ is the droplet Reynolds number. Moreover, in Eq. (7), $\dot{Q}_c = h_c A_d (T - T_d)$ denotes the convective heat transfer between two phases. Here $A_d$ is the droplet surface area and $h_c$ is the convective heat transfer coefficient, estimated using Ranz and Marshall correlations [40] through the modified Nusselt number, i.e. $\widetilde{Nu} = 2 + [(1 + Re_d Pr)^{1/3} \max(1, Re_d)^{0.077} - 1]/F(B_T)$. The gas Prandtl number $Pr$ is 1.0 and $B_T$ is the Spalding heat transfer number. Furthermore, $\dot{Q}_{lat}$ in Eq. (7) accounts for the heat transfer caused by the latent heat of droplet evaporation.

The influences of fuel droplets on the gas phase are realized with Particle-source-in-cell (PSI-CELL) method [41], through the source/sink terms of Eqs. (1)–(4):

$$S_m = \frac{1}{V_c} \sum_1^{N_p} n_p \dot{m}_d, \tag{9}$$

$$\mathbf{S}_F = -\frac{1}{V_c} \sum_1^{N_p} n_p (-\dot{m}_d \mathbf{u}_d + \mathbf{F}_d), \tag{10}$$

$$S_e = -\frac{1}{V_c}\sum_1^{N_p} n_p(-\dot{m}_d h_v + \dot{Q}_c), \quad (11)$$

$$S_{Y_m} = \begin{cases} S_m & \text{for the liquid fuel species,} \\ 0 & \text{for other species.} \end{cases} \quad (12)$$

Here $V_c$ is the CFD cell volume, $N_p$ is the parcel number in one cell, $n_p$ is the droplet number in one parcel, and $h_v$ is vapor enthalpy at the droplet temperature. The work by the droplet hydrodynamic force is not considered in Eq. (11) since it is of secondary importance in detonations with dilute and fine liquid droplets [42].

*2.2 Numerical method*

The governing equations of gas and liquid phase are solved by a two-phase compressible reacting flow code, *RYrhoCentralFoam* [23, 43], developed from OpenFOAM. *RYrhoCentralFoam* has been extensively validated with a wide range of benchmark problems against experimental or theoretical data [44], e.g., shock capturing, shock-flame interaction, droplet evaporation, as well as interphase momentum and heat exchanges. It has been used for various supersonic combustion and detonation problems [13, 25, 45-47].

For the gas phase, cell-centered finite volume method is used. Second-order backward scheme is employed for temporal discretization and the time step is about $5\times10^{-10}$ s. A MUSCL-type upwind-central scheme [48] with van Leer limiter is used for reconstruction of the convective fluxes in momentum equations, i.e. Eq. (2). The second-order central differencing scheme is applied for the diffusion terms in Eqs. (2)−(4). The chemical reaction source terms in Eqs. (3) and (4) are integrated with a Euler implicit method. The accuracy and efficiency of this integration method has been confirmed through comparing with other chemistry integration approaches [44, 49].

Two-step reactions for *n*-heptane combustion are considered, including 6 species (*n*-$C_7H_{16}$, $O_2$, CO, $CO_2$, $H_2O$ and $N_2$) [50, 51]. The kinetic parameters are listed in Table 1. It has been validated against a skeletal mechanism (44 species and 112 reactions) in our previous study [52] and the results show that it can correctly reproduce the detonation propagation speed, pressure and temperature at both von Neumann and Chapman−Jouguet (C-J) points in the ZND (Zeldovich−von Neumann−Döring) structures [24]. The two-step chemistry is sufficient for this study, since revealing detailed gas chemistry is not one of our objectives; instead, we are more interested in, e.g., propagation, structure, and gas-liquid interactions of liquid fueled detonations.

Table 1. Chemical mechanism for *n*-$C_7H_{16}$ combustion (units in cm-sec-mole-cal-Kelvin). $A$ is the pre-exponential factor, $n$ is the temperature exponent, $E_a$ is the activation energy, $a$ and $b$ are the fuel and oxidizer reaction orders, respectively.

|    | Reaction | $A$ | $n$ | $E_a$ | $a$ | $b$ |
|----|----------|-----|-----|-------|-----|-----|
| I  | 2*n*-$C_7H_{16}$ + 15$O_2$ $\Rightarrow$ 14CO + 16$H_2O$ | $6.3 \times 10^{11}$ | 0.0 | 30,000.0 | 0.25 | 1.5 |
| II | 2CO+$O_2$ $\Leftrightarrow$ 2$CO_2$ | $4.5 \times 10^{10}$ | 0.0 | 20,000.0 | 1.0 | 0.5 |

For the liquid phase, the droplets are tracked based on their barycentric coordinates. The droplet equations, i.e., Eqs. (5)−(7), are solved using first-order implicit Euler method. Validations of the sub-models (e.g., evaporation and drag force) in the droplet phase are performed in our recent work [44], and good accuracies are demonstrated. Details of the numerical method in *RYrhoCentralFoam* can be found in Refs. [44, 45, 53, 54].

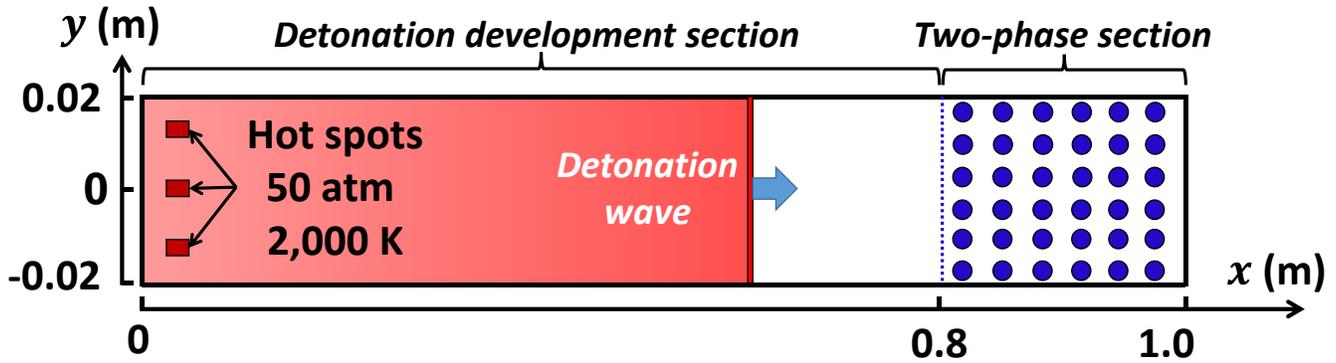

Fig. 1 Computational domain and initial *n*-heptane droplet distribution. The background gas is *n*-heptane vapour / air mixture with equivalence ratio of 0.6.

## 3. Physical problem and numerical implementation

Detonation wave (DW) propagation in heterogeneous *n*-heptane vapor / droplet / air mixtures is studied based on a two-dimensional (2D) configuration. 2D detonation simulations can well reproduce key detonation characteristics from experiments or three-dimensional simulations [55-57], e.g., C−J speed, transverse wave speed, and cell size. The computational domain is shown in Fig. 1. The length (*x*-direction) and width (*y*-direction) are 1,000 mm and 40 mm, respectively. The left and right boundaries are non-reflective, whereas zero gradient conditions are enforced at the upper and lower boundaries. The initial pressure and temperature of the background gas (i.e., *n*-heptane vapor / air) are 0.5 atm and 300 K, respectively. As marked in Fig. 1, three spots with high temperature (2,000 K) and pressure (50 atm) are used near the left boundary to initiate a detonation wave.

The domain includes: (1) gaseous detonation development section with a size of [0, 0.8 m] × [-0.02 m, 0.02 m]; (2) two-phase section of [0.8 m, 1.0 m] × [-0.02 m, 0.02 m]. The first section is sufficiently long to minimize the detonation overdrive effects and a freely propagating detonation wave can be achieved before the next section. Moreover, the development section is filled with premixed *n*-heptane vapor/air mixture and the equivalence ratio (ER) is 0.6.

The two-phase section is initialized with *n*-heptane vapor / droplet / air mixtures, and the effects of liquid droplet ER and initial droplet size on the incident DW will be studied. In all our simulations, *n*-heptane vapor ER (termed as vapor ER hereafter), $\phi_g$, is 0.6 in the two-phase section, same as that of the detonation development section. Under this vapor ER, the mixture is unsaturated and hence the droplets ahead of the DW can continuously vaporize, leading to an increased vapor concentration compared to the initial gas. Based on *a posterior* examination of our results, the actual vapor ER before the DW ranges from about 0.62 (large droplets) to 0.8 (small droplet cases).

Moreover, the liquid fuel ER (or liquid ER for short), $\phi_l$, is defined as the mass ratio of the droplets to the oxidizer normalized by the fuel-oxidizer ratio under stoichiometric condition. The liquid ER $\phi_l$ varies from 0 to 3.4 in our studies, and the initial droplet diameters are $2.5-10$ μm. The initial temperature, material density, and specific heat capacity of *n*-heptane droplets are 300 K, 680 kg/m$^3$, and 2,246 J/kg/K, respectively.

With the above range of droplet size and liquid ER, the *n*-heptane droplet number varies between 0.04 million and 21.1 million in our simulations. Since the computational parcel method is used, the droplet number in each parcel, varies from 1 to 8 depending on specific cases, leading to different droplet resolutions, i.e., the total parcel number of 0.04 million$-$2.64 million. Further analysis shows that our results are not sensitivity to the droplet resolutions used in the simulations (see details in the supplementary document).

The domain in Fig. 1 is discretized with 4,000,000 uniform Cartesian cells, and the cell size is 100×100 μm$^2$. This size is much larger than the droplet diameters, which is an intrinsic requirement of the point-force assumption in the Eulerian−Lagrangian approach. This ensures that the gas phase quantities near the droplet surfaces can be well approximated using the interpolated ones at the location of the sub-grid droplet [58]. Furthermore, mesh sensitivity analysis is performed for the two-phase section, through halving the cell size to 50 μm. The details can be found in the supplementary document, and it is found that both average detonation cell size and evolutions of the DW speed predicted with the two meshes are generally close.

## 4. Results and discussion

*4.1 Spray detonation propagation speed*

Figure 2 shows the average detonation speed as a function of liquid ER when three initial droplet diameters are considered changes. The average speed is estimated from the length (i.e., 0.8−1.0 m) of the two-phase section divided by the time with which the DW crosses it. For a given $d_0$ (e.g., 2.5 μm), the average speed $D$ first increases compared to that of the background gas (i.e., pure gas in Fig. 2) and then decreases with the liquid ER $\phi_l$. The increase is probably caused by more fuel vapor available from droplet evaporation when $\phi_l$ increases, whilst the decrease is due to the enhanced energy absorption from the gas phase due to droplet heating and phase change. Therefore, the competition between the kinetic effect ($D$ ↑) and thermal effect ($D$ ↓) exists.

The abovementioned tendency is observed for all diameters in Fig. 2. Such dependence of the DW speed on liquid ER is featured by an optimal liquid ER, which corresponds to a maximum DW speed. For the smaller droplets (2.5 and 5 μm), the optimal $\phi_l$ is around 0.4, which corresponds to an approximately stoichiometric condition of the two-phase mixture (since $\phi_g = 0.6$). Nonetheless, for coarse sprays (e.g., 10 μm), the optimal value lies in the fuel-rich range due to relatively slower droplet evaporation. Moreover, from liquid fuel utilization point of view, to achieve the same speed enhancement (e.g., 10%) relative to the background detonable gas, less fuels are needed if they are well atomized.

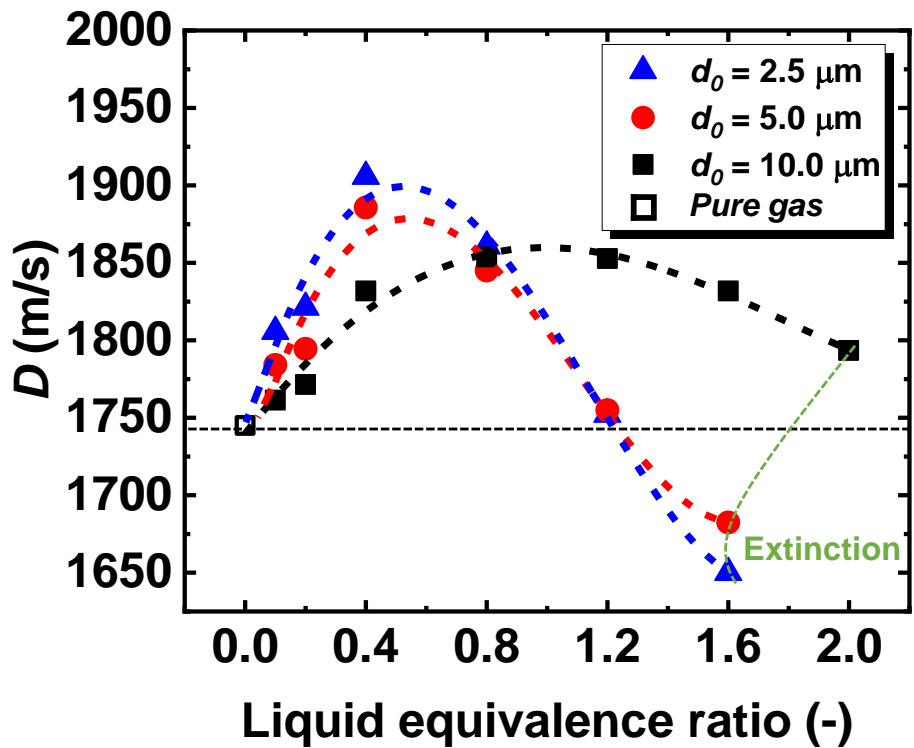

Fig. 2 Average detonation speed as a function of liquid equivalence ratio with different initial droplet diameters. Green line: extinction boundary line. Black line: average detonation speed of the droplet-free case ($\phi_l = 0$).

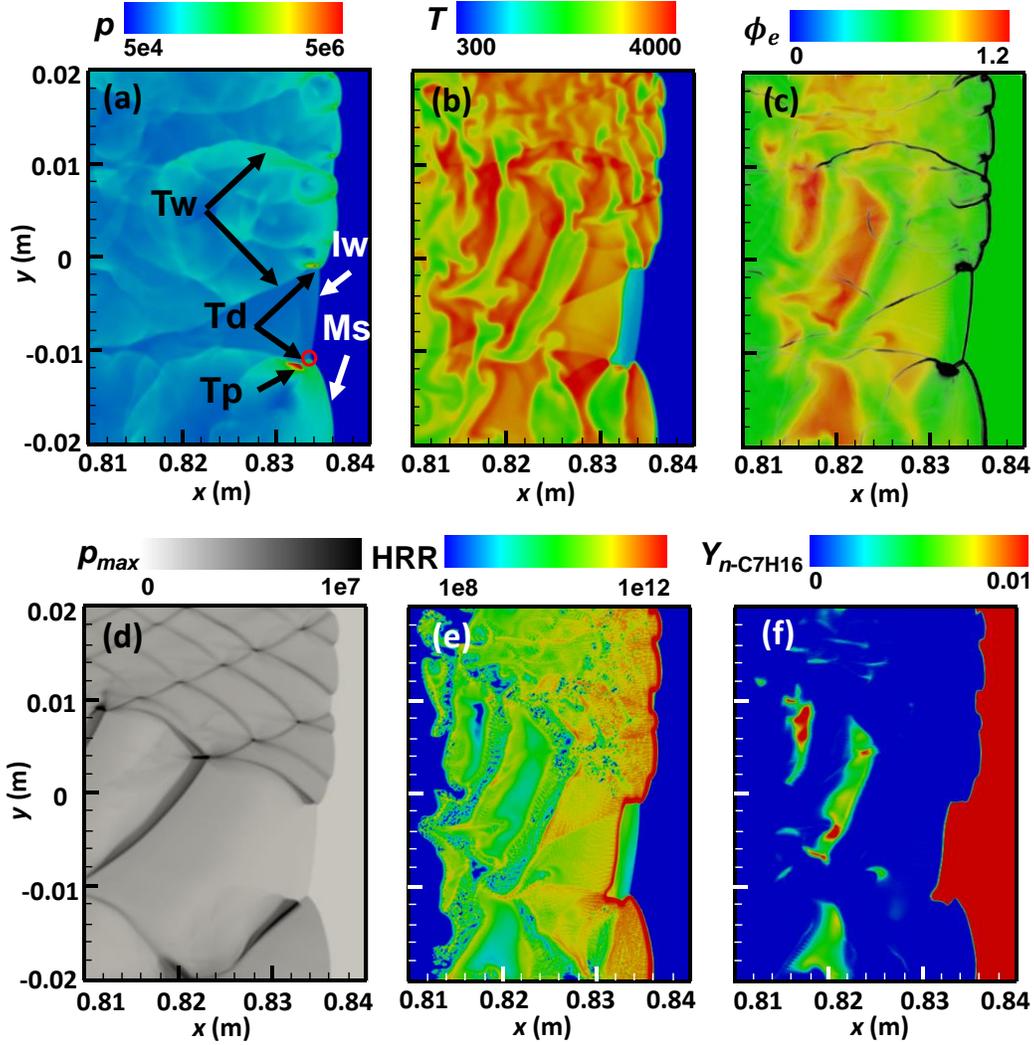

Fig. 3 Distributions of (a) pressure (in Pa), (b) gas temperature (K), (c) effective equivalence ratio, (d) peak pressure trajectory (Pa), (e) heat release rate (J/m$^3$/s), and (f) $n$-heptane vapour mass fraction. $\phi_l$ = 0.4 and $d_0$ = 10 μm. Tw: transverse wave; Td: transverse detonation; Tp: triple point; Ms: Mach stem; Iw: incident wave. Background contours of Fig. 3(c): pressure gradient magnitude.

Nonetheless, when $\phi_l$ is beyond a critical value, the average DW speed $D$ would be reduced to less than that of the background gas. This indicates that addition of the $n$-heptane sprays is not beneficial for enhancing detonation propagation on average. This critical liquid ER increases with the droplet size: it is 1.2 for $d_0$ = 2.5 and 5.0 μm, but for $d_0$ = 10 μm the speed enhancement can be achieved with a larger liquid ER. Therefore, among the three diameters, the 10 μm droplets (2.5 and 5.0 μm) have the relatively wide (narrow) liquid ER range for the DW propagation speed enhancement relative to the background gas. Further increasing the liquid ER causes detonation extinction (i.e., decoupling of leading shock and reaction front) in the heterogeneous mixture; see the *extinction* line in Fig. 2. For instance, with $d_0$ = 2.5 and 5 μm, successful detonation transmission in the two-phase section can be only achieved when $\phi_l$ < 1.6. However, for $d_0$ = 10 μm, it becomes $\phi_l$ < 2.0. This is because higher loading of finer droplets may result in richer composition and stronger heat absorption at the detonation front, which weakens the detonation significantly and reduces the propagating speed.

*4.2 Gas and droplet behaviours in spray detonation*

Figure 3 shows the distributions of gas quantities when the DW propagates in the two-phase section, including pressure, temperature, effective ER, peak pressure trajectory, heat release rate (HRR), and *n*-heptane vapor mass fraction. Here $\phi_g$ = 0.6, $\phi_l$ = 0.4 and $d_0$ = 10 μm. It is shown in Fig. 3(a) that the frontal structures of liquid fueled detonation, such as Mach stem, incident wave, transverse wave, and triple point, are well captured. Evident from Fig. 3(b) are the striped distributions of gas temperature, which is a peculiar feature in spray detonation and is absent in the gaseous detonation [32, 59-61]. It results from the interactions between triple point (Tp in Fig. 3a) and droplets in the two-phase medium and will be further discussed in Section 4.4. Moreover, short transverse detonations (Td in Fig. 3a) can be found near the triple points, which interacts with the local droplets and consumes the fuel vapor behind the incident wave. They are manifested by strong HRR (> $10^{12}$ J/m$^3$/s), high pressure gradient, and thick trajectories of peak pressures.

Figure 3(c) shows the distribution of the effective ER ($\phi_e$) in the gas phase, calculated from an element-based definition [62], i.e., the ratio of required stoichiometric oxygen atoms to the available oxygen atoms

$$\phi_e = \frac{4n_c + n_H}{2n_O}, \tag{13}$$

where $n_c$, $n_H$ and $n_O$ are the number of carbon, hydrogen and oxygen atoms, respectively. The stoichiometry corresponds to $\phi_e$ = 1. The reader is reminded that the effective equivalence ratio is also defined in the burned products, although we are more interested in the undetonated gas. It is seen that the effective ER near the leading shock is 0.63, only slightly higher than the initial vapor ER, which indicates that the droplet evaporation (hence vapor addition) are limited in this case. Instead, in the detonated area, some pockets with higher effective ER are present, resulting from the local droplet accumulation. Moreover, distributed deflagration in evaporating fuel sprays (identified with HRR < $1\times10^{11}$ J/m$^3$/s) proceeds behind the detonation front (see Fig. 3e). This is confirmed by very low fuel vapor mass fraction in most post-detonation area; see Fig. 3(f). Moreover, one can see from Fig. 3(f) that there are some *n*-heptane vapor pockets behind the detonation wave, which may be due to local droplet evaporation since they correspond to the high effective ER locations.

Figure 4 shows the droplet volume fraction, temperature, and evaporation rate, which correspond to the results in Fig. 3. Here the droplet volume fraction $\alpha_d$ is calculated from the ratio of total particle volume in one CFD cell to the cell volume. In OpenFOAM, one cell is used along the third direction (*z* direction in our studies) in two-dimensional studies and hence therefore the volume is calculated based on the quasi-two-dimensional cells. As shown from Fig. 4(a), along the trajectories of the triple points (e.g., directed by red arrows), $\alpha_d$ is much lower than that in the surrounding areas. This leads to a cellularized distribution of droplets, which is not observed from the recent Eulerian−Eulerian modelling of *n*-heptane spray detonations [32]. This may be because the Eulerian−Eulerian method homogenizes the localized gas / droplet properties, thereby failing to predict individual droplet behaviors and gas−droplet interactions [33, 63]. This is also reflected from their almost uniform droplet temperature in the post-detonation area [32]. From our results, the droplets are quickly heated to above 500 K due to high gas temperature behind the Mach stems and transverse shocks (see Fig. 4b). Therefore, strong evaporation is initiated, as demonstrated in Fig. 4(c). On the contrary, droplet temperature rises relatively slowly behind the incident waves as seen in Fig. 4(b), and hence the droplet vaporization is much delayed compared to those behind the Mach stems.

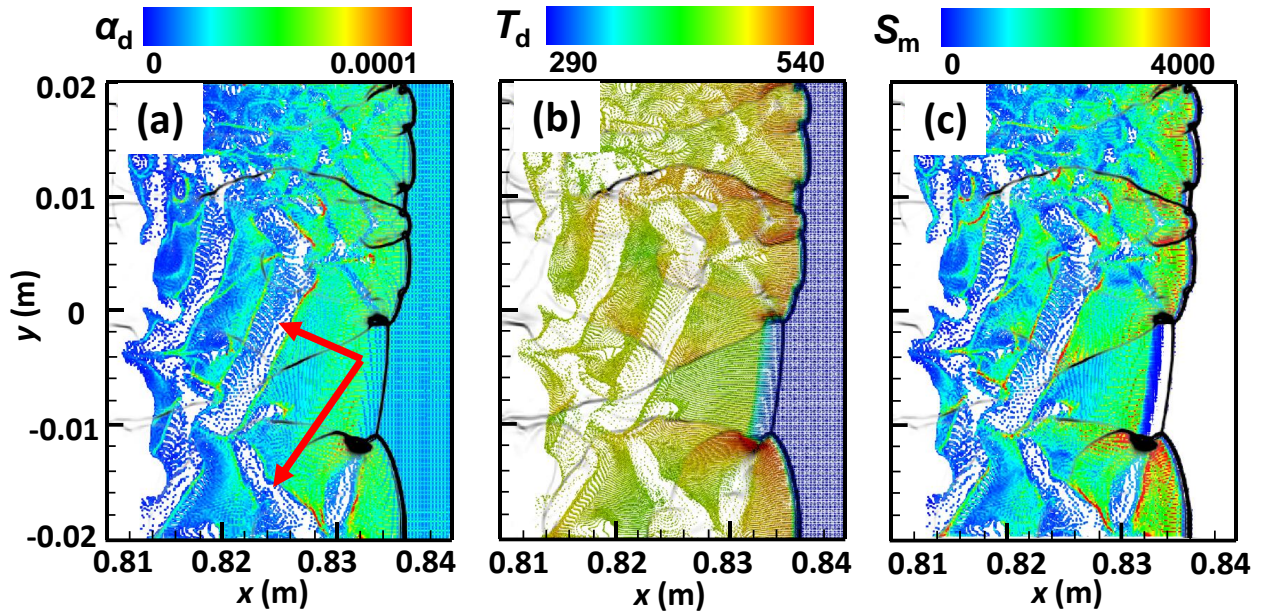

Fig. 4 Distributions of (a) droplet volume fraction, (b) droplet temperature (in K), and (c) droplet evaporation rate (kg/m$^3$/s). $\phi_l = 0.4$ and $d_0 = 10$ μm. Background contour: pressure gradient magnitude. Arrow: trajectory of the triple points.

*4.3 Reaction zone distribution*

Figure 5 compares the distributions of droplets (colored by temperature), effective ER, and HRR

with three initial droplet sizes. Note that the results of 10 μm have been presented in Figs. 3 and 4, but are added here for easy reference. Figure 6 further plots the profiles of $\phi_e$ and HRR though width-averaging (i.e., along the *y*-direction) based on the results in Fig. 5. The curves are shifted to match the leading shock front location at $x_0$ for comparison.

One can see from Fig. 5 that the droplets are quickly heated to high temperature and consumed in a short distance behind the SF when $d_0$ = 2.5 μm. The almost instant droplet gasification by detonation wave is observed by Schwer [64], in which fine JP-10 spray detonations are simulated. This results in high local $\phi_e$ in Fig. 5(b), which increases quickly and reaches a constant value, 1.1, around 0.834−0.836 m. High HRR (> 5 × $10^{12}$ J/m³/s) can be observed roughly following the SF; see Fig. 5(c). The average HRR profiles in Fig. 6 shows that pronounced increase of the effective ER is found at the HRR peak, due to fast evaporation of the droplets after the leading shock. To clarify, the secondary HRR peak in Fig. 6 arises from the combustion behind the incident wave, which again manifests the different induction lengths behind the Mach stem and incident wave. In the detonated region, approximately diamond heat release islands exist (see the red arrows in Fig. 5) and they are the relics from movement of the triple points. Similar rich pockets are also seen in Ref. [64] and it is attributed to the compression by the collision of the transverse waves. However, no details are provided in their analysis and in fact the droplets have been consumed by the detonation wave.

When the droplet size is increased to 5 and 10 μm, the droplet dispersion distance behind the DW is lengthened, as shown in Figs. 5(d) and 5(g). Meanwhile, the effective ER exhibits slower increase behind the leading SF. This can be confirmed from the averaged HRR profiles in Fig. 6. Accordingly, the effective ER near the reaction front decreases with droplet size, which can hence justfy why the HRR peak is reduced. Moreover, another feature in Fig. 6 worthy mentioning is that vaporization of fine droplets ahead of the SF causes higher $\phi_e$ in the fresh gas than vapor ER in the initial background gas (0.6). For instance, $\phi_e$ of the undetonated gas is as high as 0.8 when $d_0$ = 2.5, approximately 33% higher than the initial ER. Therefore, fine sprays are beneficial for detonative combustion not only because of fast evaporation around the DW, but also the significant pre-vaporization in the undetonated gas. The latter, however, would result in non-uniform gas composition, thereby triggering unsteady phenoemna in spray detonation [65, 66]. In detonation propulsion system (e.g., RDE), due to relatively hot carrier gas [11, 13], the pre-vaporization process probably becomes even more significant to affect the detonative combustion, particularly for highly volatile liquid fuels. It is difficult to accurately quantify the degree of *in-situ* dropley evaporation from spray detonation experiments (e.g., Ref. [67, 68]), and hence further studies are merited.

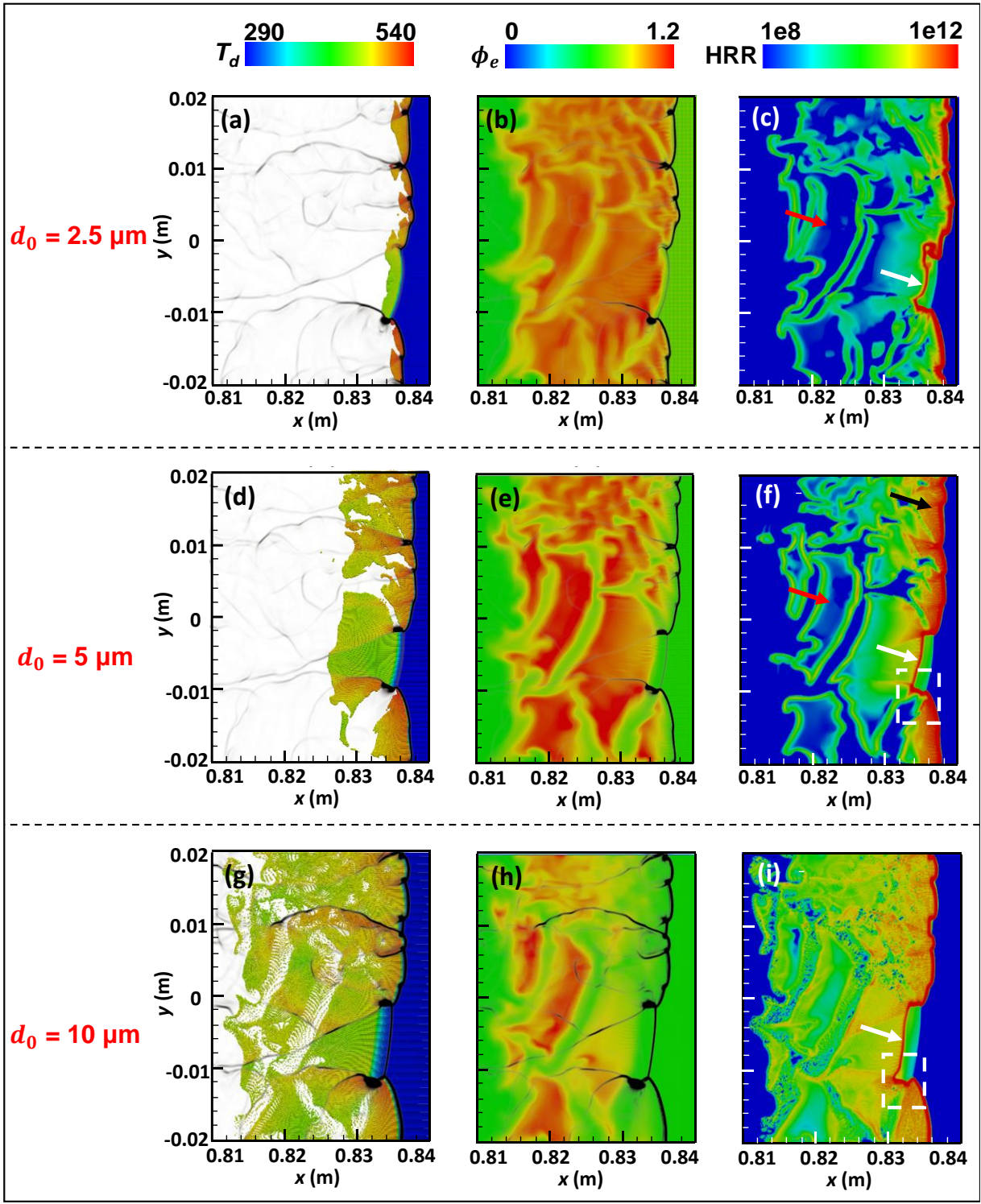

Fig. 5 Distributions of droplets coloured by temperature (in K), effective equivalence ratio, and heat release rate (J/m$^3$/s, in logarithmic scale) with different initial droplet sizes. Background contour: pressure gradient magnitude. $\phi_l = 0.4$.

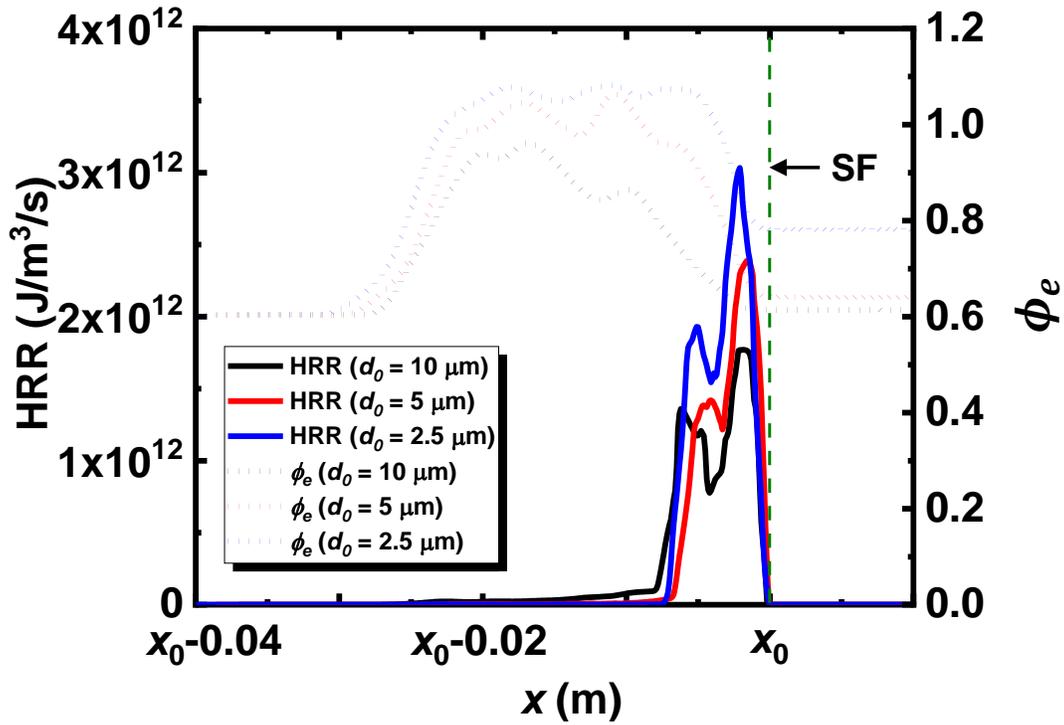

Fig. 6 Average profiles of HRR and effective equivalence ratio with different droplet sizes. $\phi_l = 0.4$.

## 4.4 Interactions between triple point and droplets

Triple point dynamics is a significant characterization of detonation frontal structure and unsteady behaviors. In spray detonations, how the triple point interacts with liquid fuel droplets will be further discussed in this section. Figure 7 shows enlarged distributions of gas and droplet quantities near the triple point, corresponding to the box of Fig. 5(f). An upward propagating transverse detonation wave is observed, extended from the triple point. Interestingly, almost no droplets are left along the propagation track (the arrow in Fig. 7), which can be justified by two reasons. Firstly, some droplets get fully vaporized subject to extremely high temperature of transverse detonation wave and triple point. For instance, the pressure and temperature of triple points can be up to $8\times10^6$ Pa and 4,000 K, respectively.

This can be confirmed by slightly increased effective ER and finite residual $O_2$ along the trajectory, as shown from the contours in Figs. 7(a) and 7(b). Secondly, strong drag force from large pressure difference (i.e., particle form drag) are significant in local flows, which makes the 5 μm droplets respond swiftly and hence move off the track of the triple point and transverse detonation. The reader is reminded that the Schiller and Naumann drag model [39] used in our studies includes the combined contributions of the shear stress and form drag. When the droplet Reynolds number is relatively high, the latter becomes dominant in the total drag force [33]. This can be seen from the *y*-component droplet velocity in Fig. 7(c) and high (low) droplet volume fraction beyond (along) the track in Fig. 7(d). As shown from Fig. 7(c), these expelled droplets still carry significant liquid fuels, shown from their finite diameters. The fuel vapor enhances the local ER to a relatively rich state (around 1.2), as revealed in Fig. 7(a). Nonetheless, inside these fuel-rich pockets, the oxygen is limited and therefore almost no reactions happen there. Instead, significant reactions are observed between the fuel-rich pockets and transverse detonation tracks (i.e., in the vicinity of fuel pockets), where some oxygen is left due to the fuel-lean combustion, which this well justifies the peculiar reaction zone distributions in Fig. 5.

As discussed in Section 4.3, when the droplets are relatively coarse, e.g., 10 μm, different droplet behaviors can be observed along the triple point trajectories. We also look at the details from the dashed box in Fig. 5(i), and the analysis is presented in supplementary document. In general, increased amount of droplets can be still observed behind the transverse detonation wave compared to that in Fig. 7. This is due to smaller evaporation rate and/or longer momentum response timescale of the larger droplets, which can also be corroborated by lower effective ER and considerable residual oxygen along the triple point / transverse detonation trajectories. This leads to relatively connected distributions of deflagration heat release in the post-detonation area, as observed from Fig. 5(i).

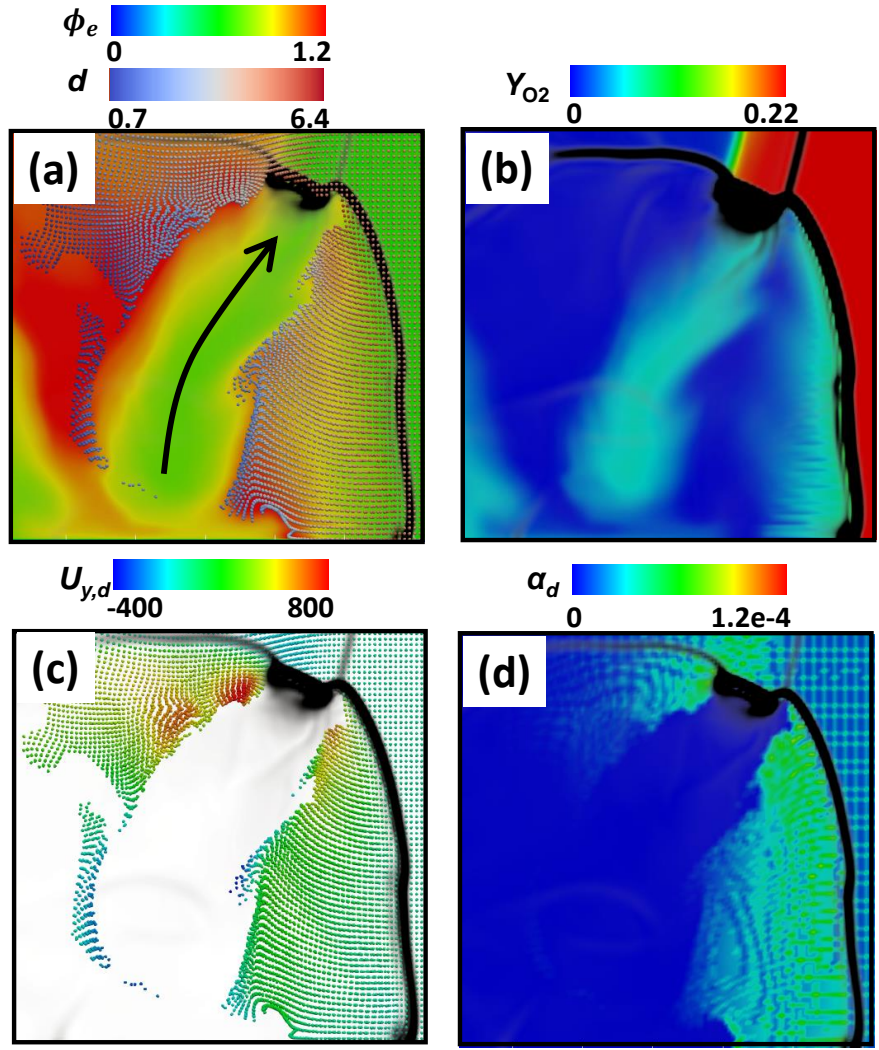

Fig. 7 Interactions between triple point and droplets: (a) fuel droplets coloured by droplet size with background of effective ER; (b) $O_2$ mass fraction; (c) distributions of liquid fuel droplets coloured by $y$-component droplet velocity; (d) contours of droplet volume fraction. $d_0 = 5$ μm and $\phi_l = 0.4$.

*4.5 Detonated fuel fraction*

The relations between combustion HRR and effective ER in the foregoing three cases are further quantified in the diagram of effective ERs versus HRR, see Fig. 8. Here HRR $< 1 \times 10^8$ J/m$^3$/s scatters are excluded, and the data are collected from eight independent instants in each case. The scatters are colored by the gas temperature. Apparently, different scatter distributions are observed when the initial droplet size changes. Specifically, when $d_0$ = 2.5 μm, the effective ER mainly ranges from 0.9 to 1.2, and most of them cluster around the stoichiometry. The detonative combustion with higher HRR happens under near-stoichiometric gas condition. The peak HRR is up to $10^{14}$ J/m$^3$/s, with the gas temperature of approximately 3,500 K. To re-iterate, the background gas ER is 0.6, and therefore evaporation of the 2.5 μm droplets significantly contributes towards the effective ER at the DW.

When $d_0$ = 5 μm, for relatively high HRR ($> 3 \times 10^{12}$ J/m$^3$/s), a large amount of scatters are featured by $\phi_e \approx 0.6$, indicating that droplet evaporation has marginal effects on the detonation, which is different from that in Fig. 8(a). Meanwhile, the range of effective ERs for deflagration becomes broader, i.e., 0.6−1.25. When $d_0$ = 10 μm, the effective ER is almost the background gas ER, 0.6, when HRR is higher than $10^{12}$ J/m$^3$/s. This manifests that the droplets exhibit limited contributions to detonative combustion in the above two cases due to limited vaporization at detonation front, which is also demonstrated by the low $\phi_e$ near the detonation front in Figs. 3(c) and 6. One can also see from Figs. 8(b) and 8(c) that high temperatures (above 4,000 K) are observed from the deflagrative combustion in the post-detonation area. For practical detonation engines, such high temperatures would constitute severe challenges, e.g., on the combustor thermal management and pollutant reduction [69, 70].

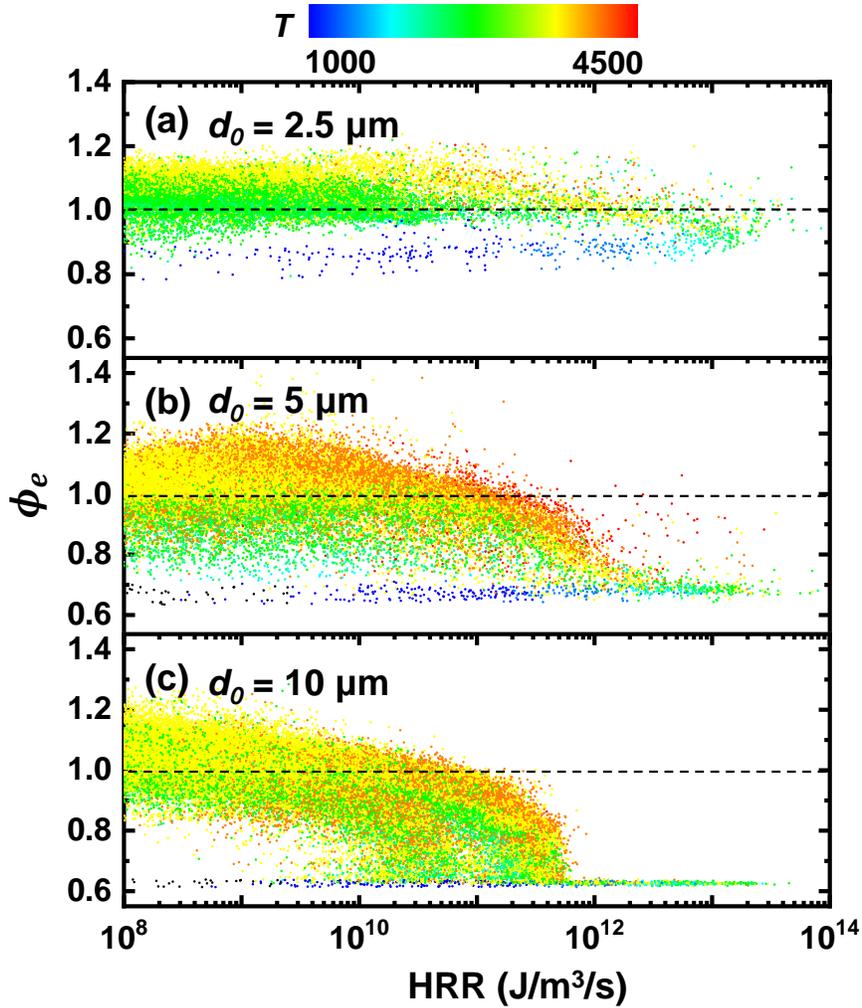

Fig. 8 Effective equivalence ratio versus heat release rate with three initial droplet diameters: (a) 2.5, (b) 5, and (c) 10 μm. The symbols are coloured by gas temperature (in K). $\phi_l = 0.4$. Dashed line: $\phi_e = 1.0$.

Contributions of the liquid fuels to detonative combustion can be quantified by detonated fuel fraction ($f_{det}$) [23]. It is defined as

$$f_{det} = \frac{ROC_{det}}{ROC_{det}+ROC_{def}}, \qquad (14)$$

where $ROC_{def}$ is the $n$-$C_7H_{16}$ consumption rate (in kg/s) by deflagration, whilst $ROC_{det}$ is the $n$-$C_7H_{16}$ consumption rates by detonation. Figure 9 shows the variations of time-averaging detonation fraction $\langle f_{det} \rangle$, $ROC_{def}$, and $ROC_{det}$ with different initial droplet sizes and liquid ERs. They are compiled from about 200 instants when the DW crosses the two-phase section. Note that the extinction cases (with relatively high liquid ER) are excluded in Fig. 9. Purely gas case ($\phi_g$ = 0.6, $\phi_l$ = 0) is also added for comparison. Besides, in order to distinguish the fuel *n*-heptane from the *in-situ* droplet evaporation and pre-vaporized *n*-heptane, we incorporate one additional reaction in the mechanism, i.e., 2*n*-$C_7H_{16}$ (v) + 15$O_2$ ⇒ 14CO + 16$H_2O$. The kinetic parameters are consistent with those of the reaction I in Table 1. The mass fractions of the *in-situ* vaporized *n*-heptane from three droplet diameters are displayed in Fig. 10. Note that the pre-vaporized *n*-$C_7H_{16}$ are fully consumed by the propagating DW and hence would not be included in Fig. 10 for brevity.

It is found from Fig. 9 that the droplet size has significant effects on the dependence of average detonated fuel fraction on liquid ER. Specifically, when $d_0$= 2.5 μm, $\langle f_{det} \rangle$ firstly increases and then decreases as $\phi_l$ increases, and a maximum value exists at $\phi_l$ = 0.6. This non-monotonicity is also observed in recent RDE modelling considering liquid *n*-heptane sprays [71]. Accordingly, in this case, $ROC_{def}$ ($ROC_{det}$) exhibits the opposite (similar) variations with the liquid ER, as seen in Figs. 9(b) and 9(c). This implies that the detonated fuel consumption rate has a dominant influence on the change of detonation fraction with fine droplets, e.g., $d_0$ = 2.5 μm. Because of their high evaporation rate, considerable vapor from the *in situ* vaporization (*n*-$C_7H_{16}$ v) can be seen immediately behind SF as shown in Fig. 10(a).

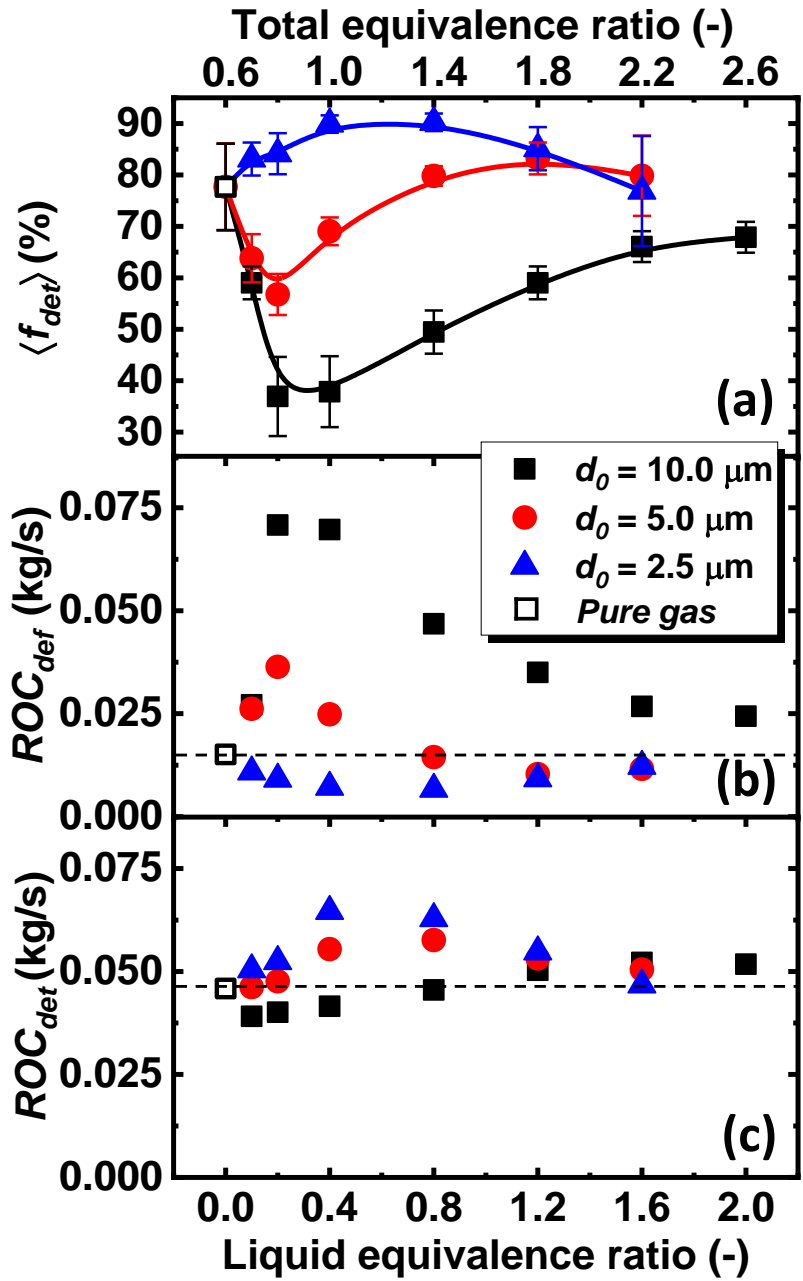

Fig. 9 Change of (a) detonated fuel fraction, (b) deflagrated fuel consumption rate, and (c) detonated fuel consumption rate with liquid equivalence ratio. Error bar: standard deviation. Dashed line: deflagrated or detonated fuel consumption rate in pure gas case.

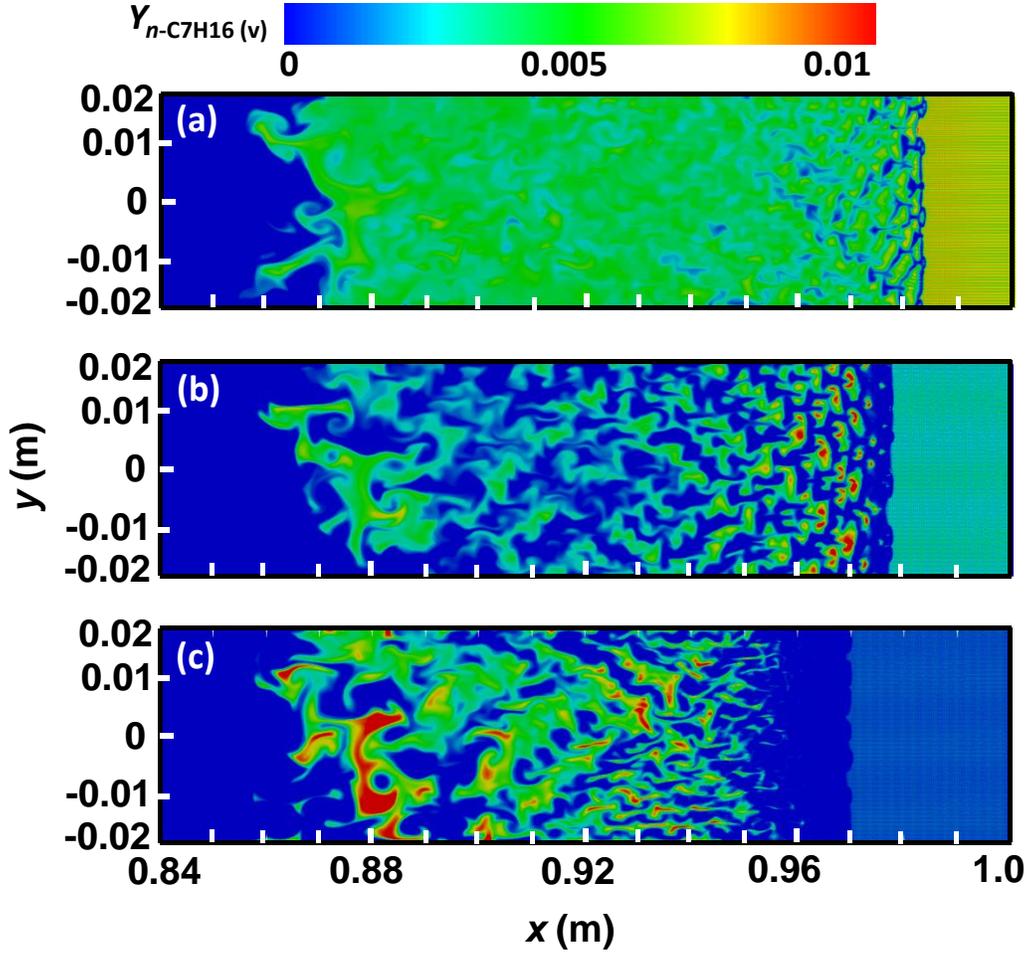

Fig. 10 Distributions of mass fraction of *n*-heptane fuel vapour from *in-situ* evaporation with different droplet diameters: (a) 2.5, (b) 5, and (c) 10 μm. $\phi_l = 0.4$.

Moreover, a V-shaped dependence of the detonated fuel fraction $\langle f_{det} \rangle$ on the liquid ER is observed for $d_0 =$ 5 and 10 μm. When small amount of liquid fuel droplets is loaded (e.g., $\phi_l = 0.1$), $\langle f_{det} \rangle$ decreases. In the meantime, $ROC_{def}$ substantially increases when $d_0 =$ 5 μm (0.015 kg/s → 0.026 kg/s) and $d_0 =$ 10 μm (0.015 kg/s → 0.027 kg/s) as shown in Fig. 9(b). This is because the surviving droplets vaporize behind the detonation front and fuel vapor is deflagrated, which significantly increases the fraction of deflagrated *n*-heptane. This can be seen from the increased mass fraction of $n\text{-}C_7H_{16}$ (v) behind the detonation front as displayed in Figs. 10(b) and 10(c). Meanwhile, $ROC_{det}$ exhibits limited changes, because the loading of sprayed droplets is still small ($\phi_l = 0.1$) and droplet evaporation mainly occurs in the post-detonation area, see Figs. 10(a)−10(c). After $\phi_l > 0.2-0.3$, $\langle f_{det} \rangle$ increases with the liquid ER. This can be justified by the fact that the effective ER around the detonation wave gradually approaches stoichiometry and therefore the detonative combustion is intensified. Since more droplets are detonated, less droplets are deflagrated behind the detonation front. Besides, in Fig. 9(b), $ROC_{def}$ decreases rapidly because more droplets absorb heat, which leads to a relatively low temperature and fuel-rich composition behind the DW. It should be mentioned that this increased $\langle f_{det} \rangle$ is achieved at the expense of spraying more fuels. Therefore, from the perspective of practical liquid fuel utilization in detonation engines, it may not be attractive, e.g., due to possible initiation of fuel-rich combustion and strong heat absorption by sprays in the combustor (hence with unsteady detonative combustion, as will be discussed in Section 4.7).

## *4.6 Spray detonation structure*

Figure 11 shows the contours of shock-frame Mach number when the initial droplet diameters are 2.5, 5, and 10 μm, respectively. Gaseous detonation with stoichiometric *n*-heptane/air mixture is also added for comparison in Fig. 11(d). Note that the upper limit of the color bar is clipped to 1.0 to clearly visualize the subsonic regions. In fact, the Mach number before the leading shock is around 6.0, due to the supersonic DW. Nonetheless, behind the leading SF, the flows become subsonic and the subsonic regions are expanded longitudinally (hence increased distance between SF and sonic line) as the droplet diameter increases from 2.5 to 10 μm, as can be observed from the width-averaged sonic location (dashed line) in Fig. 11. This may be because the leading SF is stronger when the finer droplets are loaded, which leads to larger supersonic region behind the SF. Relative to the gaseous detonation in Fig. 11(d), the distance between the SF and sonic line is reduced (increased) when the smaller (larger) droplets are loaded, corresponding to slightly enhanced (decreased) shock intensity.

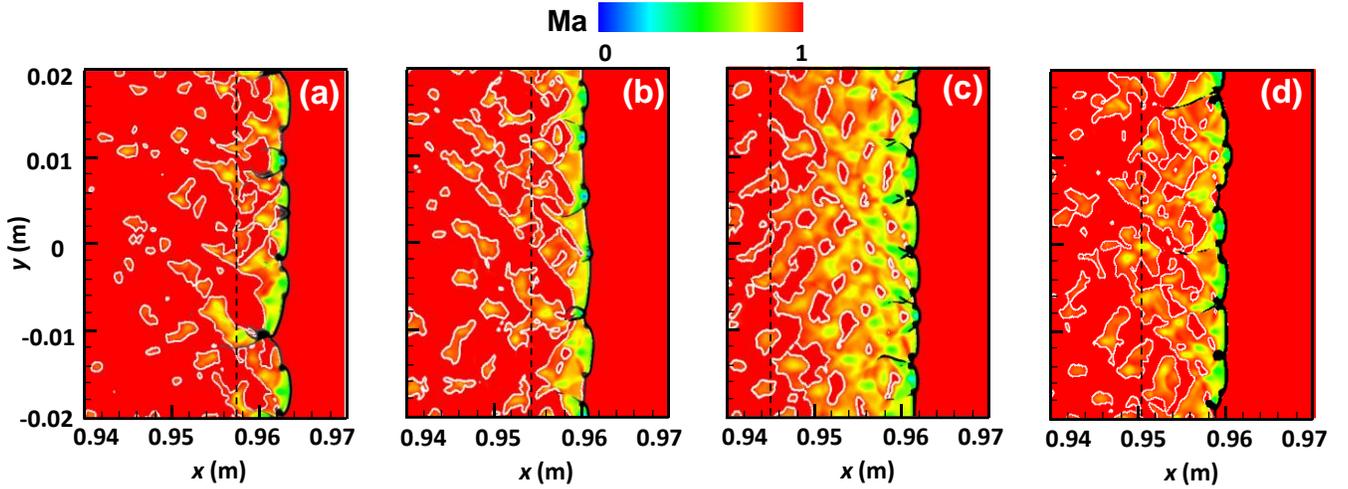

Fig. 11 Contours of shock-frame Mach number with different droplet diameters: (a) 2.5, (b) 5, (c) 10 μm, and (d) gaseous detonation. $\phi_l = 0.4$ for two-phase case, and $\phi = 1.0$ for gaseous detonation. White isoline: sonic line; dashed line: width-averaged sonic line.

In Fig. 12 we average the key gas and liquid quantities along the domain width ($y$-direction) corresponding to the same instants. The spatial profiles of droplet evaporation rate, droplet temperature and diameter, shock-frame Mach number, effective ER, gas temperature, and HRR are presented in Fig. 12. When $d_0 = 2.5$ μm in Fig. 12(a), the Mach number ahead of the DW is 6.2. Behind the leading SF, the flows continuously decelerate until the subsonic condition ($Ma < 1$) is reached at about 6 mm downstream of the SF. Beyond the subsonic zone, the flows become lightly supersonic, featuring a freely propagating detonation wave. The location where the sonic condition is first recovered from subsonic condition is termed as *sonic point* (SP, annotated in Fig. 12), which prevents downstream perturbation propagating towards the DW [72, 73].

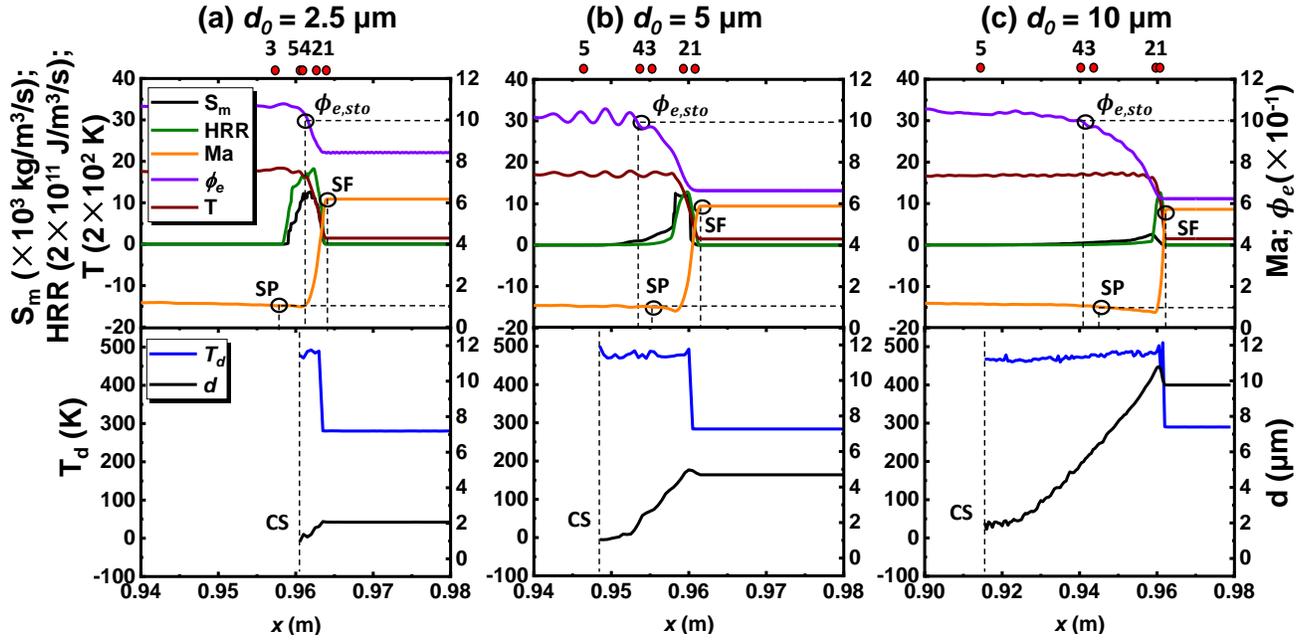

Fig. 12 Profiles of droplet evaporation rate, HRR, shock-frame Mach number, effective ER, gas temperature, droplet temperature, and droplet diameter. $\phi_l = 0.4$. 1: leading shock front; 2: reaction front; 3: sonic point; 4: stoichiometric point; 5: contact surface.

One can also see from Fig. 12(a) that the average droplet diameter before the leading SF is around 2 μm, which indicates that the droplets have initiated evaporation before the DW arrives (i.e., $x > 0.964$ m). Consequently, the effective ER $\phi_e$ of the undetonated gas is 0.8, 33% higher than the initial gas ER ($\phi_g = 0.6$). In fact, existence of the droplet pre-vaporization leads to non-uniform fuel vapor distribution before the DW, although this is affected by gas thermochemical conditions and liquid fuel volatility alike. Based on our estimations from the ZND structure of gaseous *n*-heptane detonation, the HRL is halved (from 0.48 mm to 0.21 mm) due to increased ER from 0.6 to 0.8. Furthermore, in the shocked gas the droplets are quickly heated to their saturation temperature (about 480 K). Accordingly, the droplet evaporation rate $S_m$ quickly increases and peaks at around 2 mm behind the SF. Due to addition of the fuel vapor, the effective ER increases accordingly. Moreover, droplet evaporation proceeds before the SP and the peak evaporation rate ($S_m$) lies around the RF. This again indicates the direct contributions of fuel vapor towards the local detonative combustion. In this study, the location where the local gas first becomes stoichiometric is termed as *stoichiometric point* ($\phi_{e,sto}$), which is $x = 0.9613$ m for the 2.5 μm case. The droplets are fully vaporized at $x = 0.96$ m, resulting in a *contact surface* (CS), demarcating the upstream heterogeneous (two-phase) and downstream homogeneous (gas-only) mixtures. The foregoing skeletal locations have been projected to the top $x$ axis, which includes (from right to left): leading SF, RF, stoichiometric ER, two-phase contact surface, followed by the sonic point.

The influences of droplet size on the spray detonation structure are further examined through comparing Figs. 12(a)−12(c), in which the results of 5 μm and 10 μm droplets are also shown. Likewise, their key locations are added the top $x$ axis of Figs. 12(b) and 12(c). Firstly, the shock-frame Mach number ahead of the DW decreases with the initial droplet diameter. Specifically, $Ma$ is 6.1 when $d_0 = 2.5$ μm, but reduced to 5.7 when $d_0 = 10$ μm. Low $Ma$ leads to more downstream SP relative to the leading SF, as indicated in Fig. 12 and confirmed in Fig. 11. Besides, slower evaporation of the coarse droplets results in larger distance between the stoichiometric points and leading SF. In particular, when $d_0 = 10$ μm, the stoichiometric point is even slightly beyond the SP, as demonstrated in Fig. 12(c). Besides, lower evaporation rate results in longer droplet dispersion distance in the detonated flows, characterized by longer CS−SF distance. Also, different from the 2.5 μm case, when $d_0 = 5$ and 10 μm, the CS lies well behind the SP, and the CS−SP distance increases with the droplet size. The fuel vapor from the post-SP droplets would react with the local oxidizer, which however has limited contributions towards the detonative combustion through the RF−SF coupling. Based on our calculations from Figs. 11 and 12, about 17% and 24% of total fuel vapor released from droplets are burned downstream of the SP.

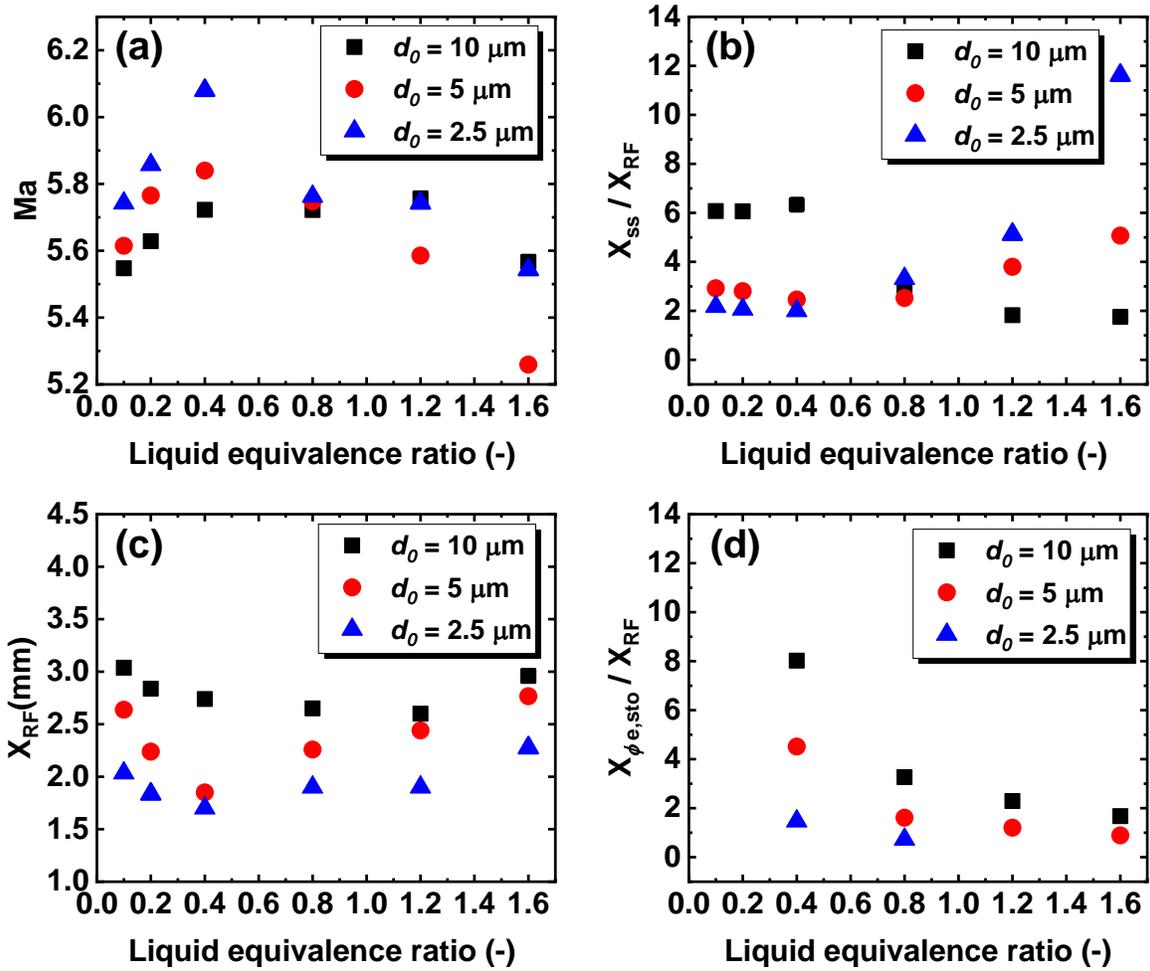

Fig. 13 Changes of key locations in detonation structure with initial droplet diameter and liquid equivalence ratio: (a) shock-frame Mach number, (b) distance from shock front to sonic point, (c) distance from shock front to reaction front $X_{RF}$, and (d) distance from shock front to stoichiometric point. Distances of SF−SP ($X_{ss}$) and SF−$\phi_{e,sto}$ ($X_{\phi e,sto}$) are normalized by distance SF−RF ($X_{RF}$).

Figure 13 summarizes the change of key location in spray detonations with initial droplet sizes and liquid ERs. The results in Fig. 13 is averaged from five instants when the DW propagates in the two-phase domain. In general, the liquid fuel properties have significant influences on spray detonation structures. The shock-frame Mach number, $Ma$, firstly increases and then decreases for a fixed droplet diameter as the liquid ER $\phi_l$ varies from 0.1 to 1.6. The increase is caused by the increased fuel vapor consumed by detonative combustion (kinetic effects), corroborated by the variation of $ROC_{det}$ in Fig. 9(c). Moreover, decreasing Mach number results from the weakened DW due to enhanced evaporative heat loss. These non-monotonicity can be observed for all diameters in Fig. 13(a). Higher $Ma$ (stronger detonation wave) generally leads to higher flow speed behind the DW and hence the subsonic flows can quickly accelerate to sonic condition in the shock frame. Thus, opposite change of the SF-SP distance can be observed from Fig. 13(b).

Figure 13(c) shows that change of the RF−SF distance, $X_{RF}$, for three diameters. $X_{RF}$ firstly decreases and then increases with $\phi_l$. Stronger leading shock can ignite the detonable gas earlier, which corresponds to reduced RF−SF distance (i.e., induction zone length). Furthermore, plotted in Fig. 13(d) is the distance from leading SF to stoichiometric point. When $\phi_l < 0.4$, the stoichiometric condition is not satisfied at any locations of the detonation structure ($\phi_t < 1$). When $\phi_l > 1.0$ and small droplets (2.5 μm), fuel-rich composition is achieved ahead of the DW due to fast *in-situ* vaporization, which can be found from the distributions of the effective ER of the $\phi_l = 1.2$ and 1.6 cases in the supplementary document. Thus, they are excluded in Fig. 13(d). In general, this distance decreases as $\phi_l$ increases for all droplet sizes. This is because more fuel vapor becomes available when liquid ER increases, and hence the location where the stoichiometric condition is reached becomes shorter behind the leading SF.

*4.7 Detonation extinction and re-initiation*

Up to this point, we have discussed the results when the detonation wave can successfully transit the two-phase section. Detonation extinction and re-initiation, featured by the SF and RF decoupling / re-coupling, are observed when the liquid ER is high. In Fig. 14, we visualize a localized extinction process along the detonation front when $d_0$ = 10 μm and $\phi_l$ = 2.4. The animation of this case can be found from the supplementary document. Note that only top half domain is illustrated and the top boundary behaves like a nonpermeable wall. Figure 15 presents the corresponding evolutions of the interphase heat transfer rate (HTR, see Eq. 11), HRR, and effective ER. Negative HTR indicates that the energy is transferred from the gas to droplet phase.

At 0.000491 s, a weakly unstable detonation front is observed with four triple points (Tp), which are numbered with #1−#4 in Fig. 14(a). Tp #1, #2, and #4 are moving downward (see the arrows) and leaves the domain later, whilst Tp #3 is upward. At this moment, average heat transfer rate is about -6.6 × $10^{11}$ J/m³/s behind the Mach stem and such high heat absorption increases the SF−RF distance. After 0.000491 s, Tp #2 collides with Tp #3, producing Tp #2' and #3' in Fig. 14(b). Movement of triple points makes the upper section of the SF lack triple point collision. Meanwhile, significant heat absorption for droplet heating continues behind the Mach stem, but limited vapor is produced locally, leading to negligible increase of the gas ER, as revealed in Fig. 15(b). Based on our results, the heat absorption would further increase when the liquid fuel loading is increased. For instance, when $d_0$ = 10 μm and $\phi_l$ = 2.8, full decoupling of the SF and RF along the DW is observed. Brief discussion for this example can be found in the supplementary document.

As the Mach stem is continuously weakened, the length of induction zone behind it evidently increases at 0.000497 s in Figs. 14(c) and 15(c). After the collision between Tp #1 and #3', the new Tp #3" moves upward, consuming the shocked gas behind the Mach stem. At 0.000505 s, Tp #3" approaches the upper boundary. Meanwhile, the induction zone length is further increased and the HRR is lower than 4 × $10^{11}$ J/m³/s (see Fig. 15c). Therefore, a localized and instantaneous extinction along the DW occurs.

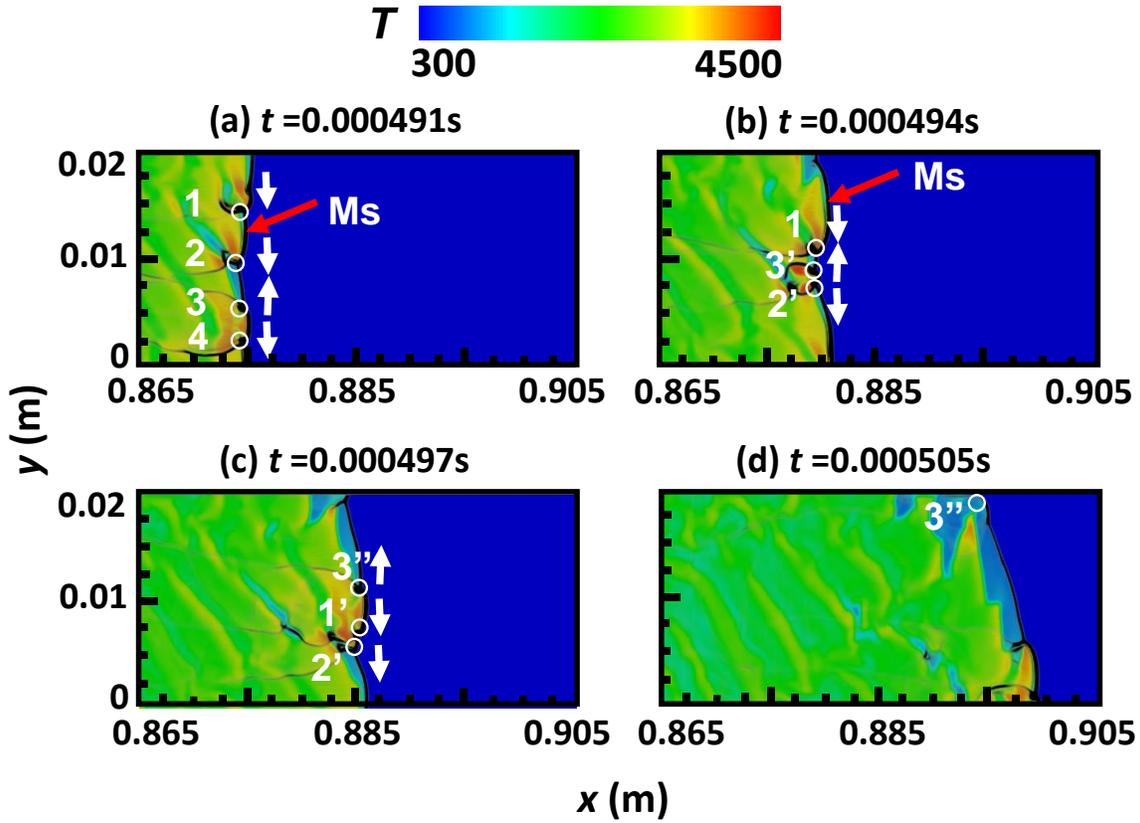

Fig. 14 Time sequence of gas temperature (in K). Pressure gradient magnitude is overlaid for visualizing the shock structure. $d_0 = 10$ μm and $\phi_l = 2.4$. Ms: Mach stem.

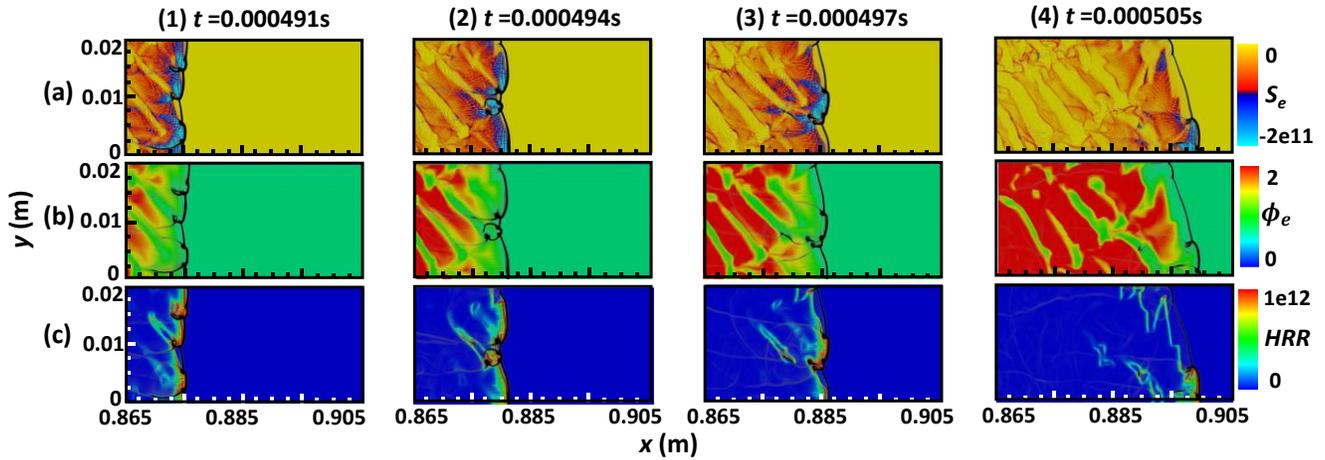

Fig. 15 Time sequence of (a) interphase heat transfer rate (in J/m$^3$/s), (b) effective ER, and (c) HRR (J/m$^3$/s). Pressure gradient magnitude is overlaid for visualizing shock structure. $d_0 = 10$ μm and $\phi_l = 2.4$.

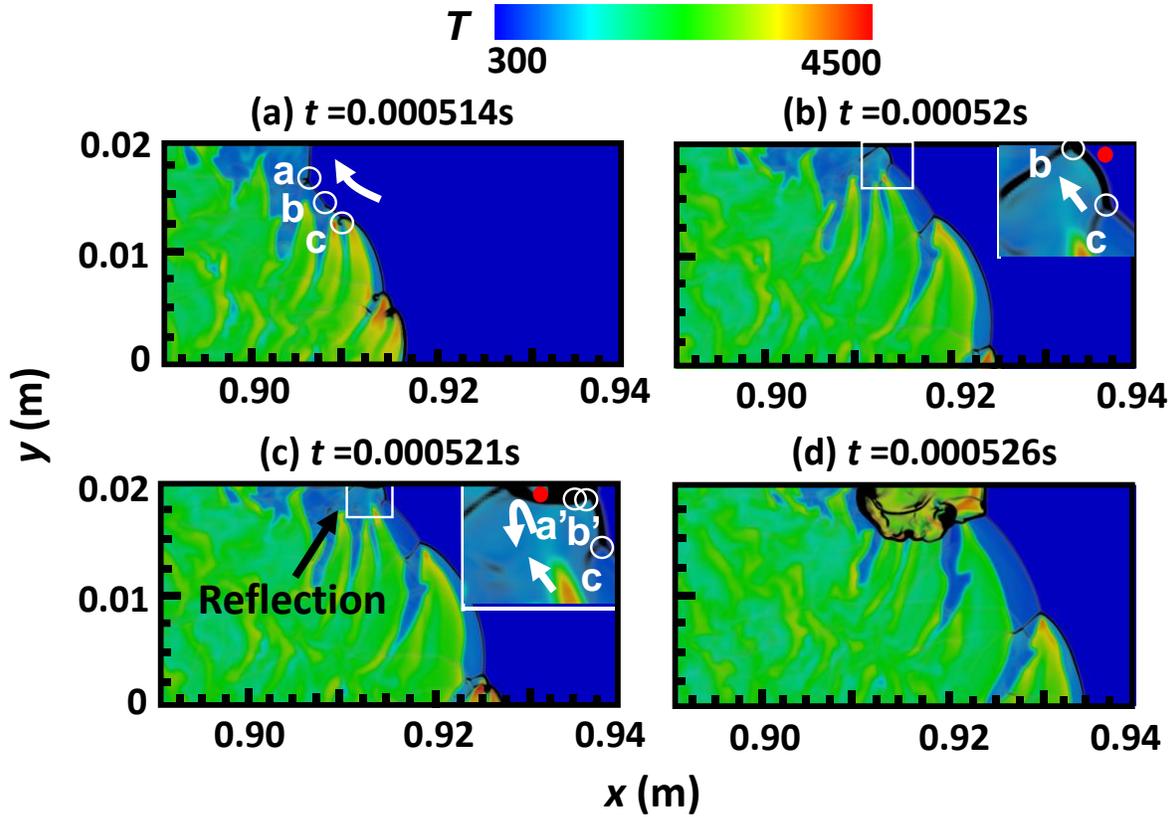

Fig. 16 Time sequence of gas temperature (in K). Pressure gradient magnitude is overlaid for visualizing shock structure. $d_0 = 10$ μm and $\phi_l = 2.4$.

After localized extinction happens, more triple points move upward in Fig. 16(a), and they arrive at the upper boundary at 0.00052 s. How the transverse shocks extended from triple points interact with the boundary and hence affect local thermochemical states can be analyzed through a probe $P$, i.e., red circle in Fig. 16. Here we plot the history of gas temperature, effective ER, pressure, and HRR at $P$ in Fig. 17. Four sequential pressure rises, labeled with (1)–(4), are observed. The first rise is caused by the Ms between Tp b and c, see Fig. 16(b). Tp a just reaches the upper boundary at this instant. Then Tp a and b reflect on the upper boundary, generating Tp a' and b' at 0.000521 s, as shown in Fig. 16(c), and the transverse shocks lead to the second pressure rise. Note that Tp a' and b' almost coalesce while approaching the probe location, and hence only one pressure rise is present. The third pressure rise results from the collision between the transverse shocks from coalescent Tp (from a' and b') and Tp c, which re-initiates strong chemical reactions characterized by high local pressure (~ 5 × $10^6$ Pa), HRR (~1 × $10^{12}$ J/m$^3$/s), and temperature (> 3,000 K). This further ignites the reactants in the lengthened induction zone and the emanated pressure waves result in the fourth pressure rise.

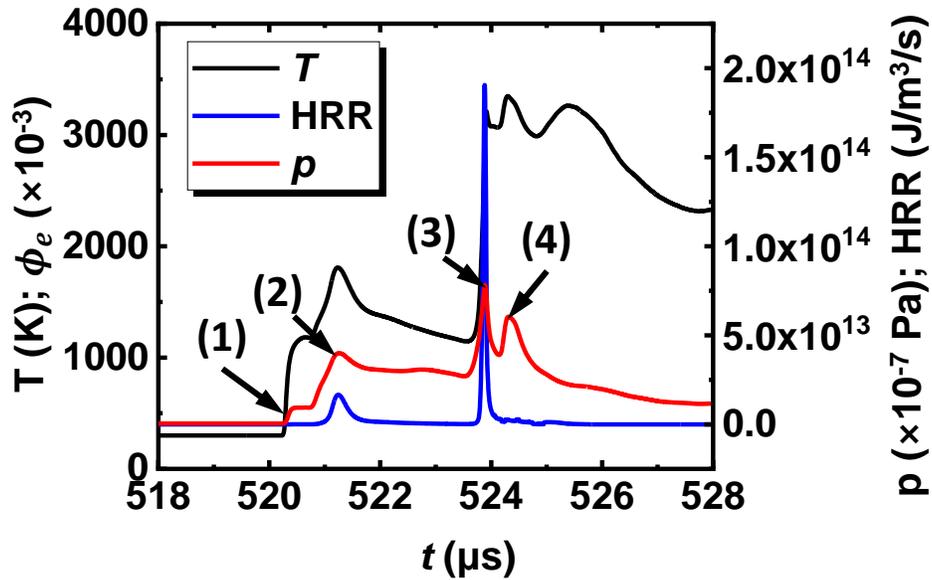

Fig. 17 Time history of gas temperature, HRR and pressure from the probe, corresponding to the red circle in Fig. 16.

## 5. Conclusion

The Eulerian–Lagrangian method with two-way gas−liquid coupling is used to study detonation propagation in two-phase mixtures of *n*-heptane vapour, droplets, and air. We consider a two-dimensional domain and two-step chemistry for *n*-heptane combustion. The background gas is *n*-heptane /air with an equivalance ratio of 0.6. The effects of droplet diameter and liquid fuel equivalence ratio on spray detonation propagation, structure, and dynamics are investigated.

Our results show that the detonation propagation speed is significantly affected by droplet diameter and liquid equivalence ratio. For a given droplet diameter, the detonation propagation speed change non-monotonically with liquid equivalence ratio. The equivalence ratio of maximum propagation speed increases with initial droplet diameter.

Droplet dispersion distance behind the detonation increases with initial droplet diameter because of slower evaporation of larger droplets. Moreover, striped gas temperature and cellularized droplet distributions can be found behind the detonation front, caused by the interactions between triple points (also transverse detonation connected with them) and liquid droplets. The local droplet accumulation further leads to isolated regions with deflagrative combustion heat release in the post-detonation area.

Detonated fuel fraction is introduced to measure how many fuels are reacted through detonative combustion. Small droplets (e.g., 2.5 μm) generally lead to high detonated fuel fraction, because they are quickly gasified and hence the vapour can be detonated. As the liquid equivalence ratio increases, the detonated fuel fraction increases first and then decreases. However, for relatively large droplets, e.g., 5 and 10 μm, detonated fuel fraction exhibits V-shaped dependence on the liquid equivalence ratio.

Spray detonation structure is also analyzed based on one-dimensionalized (averaging along the domain width) quantities of the gas and droplet phase. It is found that the spray detonation structure is significantly affected by droplet diameter and liquid equivalence ratio. Shock-framed Mach number increases first and then decreases with the liquid equivalence ratio. However, the distance from sonic plane to shock front presents almost the opposite trend. In addition, the induction length generally increases with initial droplet diameter. It first increases and then decreases with the liquid equivalence ratio, and the minimum values approximately correspond to liquid equivalence ratio of 0.4.

When the liquid equivalence ratio is sufficiently high, pronounced unsteadiness of spray detonation is observed, such as instantaneous or complete extinction. These extinctions happens due to strong energy absorption by the evaporating droplets behind the shock. Furthermore, localized detonative spot is initiated due to the compression by multiple transverse shocks.

## Acknowledgement

The simulations used the ASPIRE 1 Cluster from National Supercomputing Centre, Singapore (NSCC) and the Fugaku Cluster from High Performance Computing Infrastructure in Japan (hp210196). This work is partially supported by the open research grant (No. KFJJ20-09M) from State Key Laboratory of Explosion Science and Technology, Beijing Institute of Technology. QM is supported by National University of Singapore (Chongqing) Research Institute.